# Representational drift under spontaneous activity - self-organized criticality enhances representational reliability

Zhuda Yang[1,2], Junhao Liang[2,3], Wing Ho Yung[4], Changsong Zhou[1,2,5*]

1 Department of Physics, Hong Kong Baptist University, Kowloon Tong, Hong Kong
2 Centre for Nonlinear Studies, Institute of Computational and Theoretical Studies, Hong Kong Baptist University, Kowloon Tong, Hong Kong
3 School of Mathematics, Sun Yat-sen University, Guangzhou, P.R. China
4 Department of Neuroscience, City University of Hong Kong, Kowloon, Hong Kong SAR, China
5 Life Science Imaging Centre, Hong Kong Baptist University, Kowloon Tong, Hong Kong
* Contact author: cszhou@hkbu.edu.hk

## Abstract

Neural systems face the challenge of maintaining reliable representations amid variations from plasticity and spontaneous activity. In particular, the spontaneous dynamics in neuronal circuit is known to operate near a highly variable critical state, which intuitively contrasts with the requirement of reliable representation. It is intriguing to understand how reliable representation could be maintained or even enhanced by critical spontaneous states. We firstly examined the co-existence of the scale-free avalanche in the spontaneous activity of mouse visual cortex with restricted representational geometry manifesting representational reliability amid the representational drift with respect to the visual stimulus. To explore how critical spontaneous state influences the neural representation, we built an excitation-inhibition network with homeostatic plasticity, which self-organizes to the critical spontaneous state. This model successfully reproduced both representational drift and restricted representational geometry observed experimentally, in contrast with randomly shuffled plasticity which causes accumulated drift of representational geometry. We further showed that the self-organized critical state enhances the cross-session low-dimensional representation, comparing to the non-critical state, by restricting the synapse weight into a low variation space. Our findings suggest that spontaneous self-organized criticality serves not only as a ubiquitous property of neural systems but also as a functional mechanism for maintaining reliable information representation under continuously changing networks, providing a potential explanation how the brain maintains consistent perception and behavior despite ongoing synaptic rewiring.

## Introduction

The neural system performs complicate tasks under rich spontaneous dynamics [1,2]. Instead of being simple background noise during a task, spontaneous neural activity may be of functional significance. For instance, it may serve as a prior and generate the prediction for the incoming signals to optimize the task implementation [3,4]. However, most studies that quantify long-term neural coding have focused on stimulus-evoked states, giving comparatively little attention to how the statistics of the spontaneous dynamics shape representational structure over time. One example is the recent studies about the stimulus-evoked neural representations over a long timescale. Traditionally, one would expect that the neural representations to the same input signals tend to be identical. Recent experimental studies focusing on stable task performance over a long timescale have revealed that neuronal representations for identical input signals exhibit significant variability but also share certain reliability across sessions [5–7]. First, data from various brain regions have confirmed a robust phenomenon called representational drift, where the preferences of single neurons change over time and the neuronal firing patterns of the circuit show weak correlations across sessions [8–10]. Second and more importantly, the correlations between the representation of multiple signals are relatively reliable across sessions [8], and in human V1 such relationships can remain approximately invariant to the interval between few weeks [11], suggesting a potential mechanism for extracting reliable task information from population neural activity despite its variability on the neuronal level response to a fixed signal. Theoretical studies proposed that such representational drift may reflect the redundancy of the task-solution space in the presence of noise, where there could be multiple networks to achieve the same task or similar neural representation [12,13]. Yet, whether and how the non-trivial spontaneous dynamics actively influence this long-term representational organization remains largely untested.

The spontaneous dynamics could influence neural representation through the synaptic plasticity. Synaptic plasticity [14,15], while is crucial for adaptation and learning, inevitably modifies the network structure according to the complex and variable spontaneous firing patterns and could disrupt the established neural representations in a circuit. Thus, representational drift can be attributed to the inevitable perturbation of the synapse connections induced by synaptic plasticity from spontaneous activity. On the other hand, if spontaneous dynamics constantly trigger synaptic updates, how do neural populations retain stable relational structure? This yields a central question: does plasticity induced by spontaneous activity merely drive drift, or can it, counterintuitively, also support reliability by constraining changes to subspaces that preserve low-dimensional structure?

In particular, the spontaneous dynamics of neuronal activity exhibit non-trivial features including scale-free avalanches [16,17]. Scale-free avalanches are cascading events characterized by power-law distributions in event size and event duration. Scale-free neural avalanches are remarkably universal, having been observed across diverse brain regions and species [18–20]. Such observations align with the concept of self-organized criticality (SOC) [21,22], wherein neural systems autonomously tune their states to criticality near a phase transition. Experimental support of SOC includes in vivo studies demonstrating that mouse visual cortex gradually restores critical dynamics following monocular deprivation [23]. One potential mechanism for the SOC in the neural system is the homeostatic plasticity [15,22], which regulates the network to achieve a stable state over time despite changes in external conditions or internal activity levels. Various computational models incorporating the homeostatic plasticity rule have successfully replicated SOC phenomena [23–25]. On the other hand, critical neural dynamics is inherently highly variable and exhibits sensitive response to small perturbations, as predicted by the critical branching model [26,27] and demonstrated in experimental data [28–32], apparently at odds with the requirement for reliable responses. This further deepens the conflict between the spontaneous criticality sustained by the plasticity and long-term reliable response. The dilemma of reconciling neural response sensitivity and reliability has been addressed by recent studies revealing that the critical neural avalanches in the excitation-inhibition (E-I) balanced networks (without synaptic plasticity) exhibit reliable low-dimensional representational pattern [33] and reduced trial-to-trial variability [34] in the presence of external inputs. However, it is unknown if this model could display SOC in the presence of homeostatic plasticity and whether its response patterns to external stimuli are still reliable amid the constant network structural changes induced by the plasticity.

In summary, how the highly variable spontaneous critical dynamics induced plasticity affects the neural representation as observed in the experimental data is still an open question. Does the critical spontaneous neural dynamics only induce the representational drift, or could it also contribute to the reliability against/admit the drift? Exploring this question is of fundamental importance for understanding the functional role of spontaneous neural activity.

Here, we address this gap by jointly analyzing spontaneous and evoked neural activities in the same cortical circuits and by building an E–I spiking network model endowed with homeostatic inhibitory plasticity. We first revisited and examined experimental data where the representational drift was observed [8] and we demonstrated that its spontaneous state exhibits near-critical features. Using an E-I network incorporating homeostatic plasticity specifically at inhibitory (to excitatory) connections [35], we showed that the model self-organizes to critical spontaneous dynamics and successfully recapitulates key experimental observations from in vivo visual cortex recordings [8] - specifically the phenomena of representational drift and representational reliability manifested by restricted representational geometry of multiple signals. We found that homeostatic plasticity under the variable spontaneous activity confines the network responses into a restricted subspace, while a randomized shuffling plasticity rule disrupts both SOC and the constraint of the representational geometry. The neural dynamics operating near criticality demonstrate effective low-dimension representation compared to subcritical or supercritical regimes, which emerges from the relative stable inhibitory synapse pattern with the minimal variance near the critical state. Overall, our work provides a linkage of two actively explored fields in neuroscience, neural criticality in spontaneous states and representational drift and stability in evoked states, which are relatively independently investigated. By explicitly linking spontaneous criticality, plasticity and representational geometry, our results offer a unifying account of how reliable representation can coexist with continual synaptic change. This linkage thus provides a more unified framework towards neural circuit dynamics and information processing.

## Results

### Representational drift admits critical spontaneous activity

Neuronal circuits face a fundamental challenge: maintaining reliable representations of external signals despite ongoing synaptic rewiring under variable spontaneous activity. The typical spontaneous state is characterized by scale-free avalanches, which have been considered as evidence for the critical brain hypothesis [16]. We retested this hypothesis by evaluating neuronal avalanches from spontaneous neuronal activity recordings in the V1 region of mouse cortex (see Methods) across several sessions in in vivo experiments studying representational drift [8]. Both avalanche duration and size distributions follow power-law relationships (Fig. 1A), and the deviation from criticality coefficient (DCC), which represents the difference between observed exponent relations and those predicted by crackling noise theory (see Methods), is near zero [17,23]. These features indicate near-critical spontaneous dynamics in the mouse visual cortex across sessions, consistent with other experimental observations in the same brain region in the same species [23].

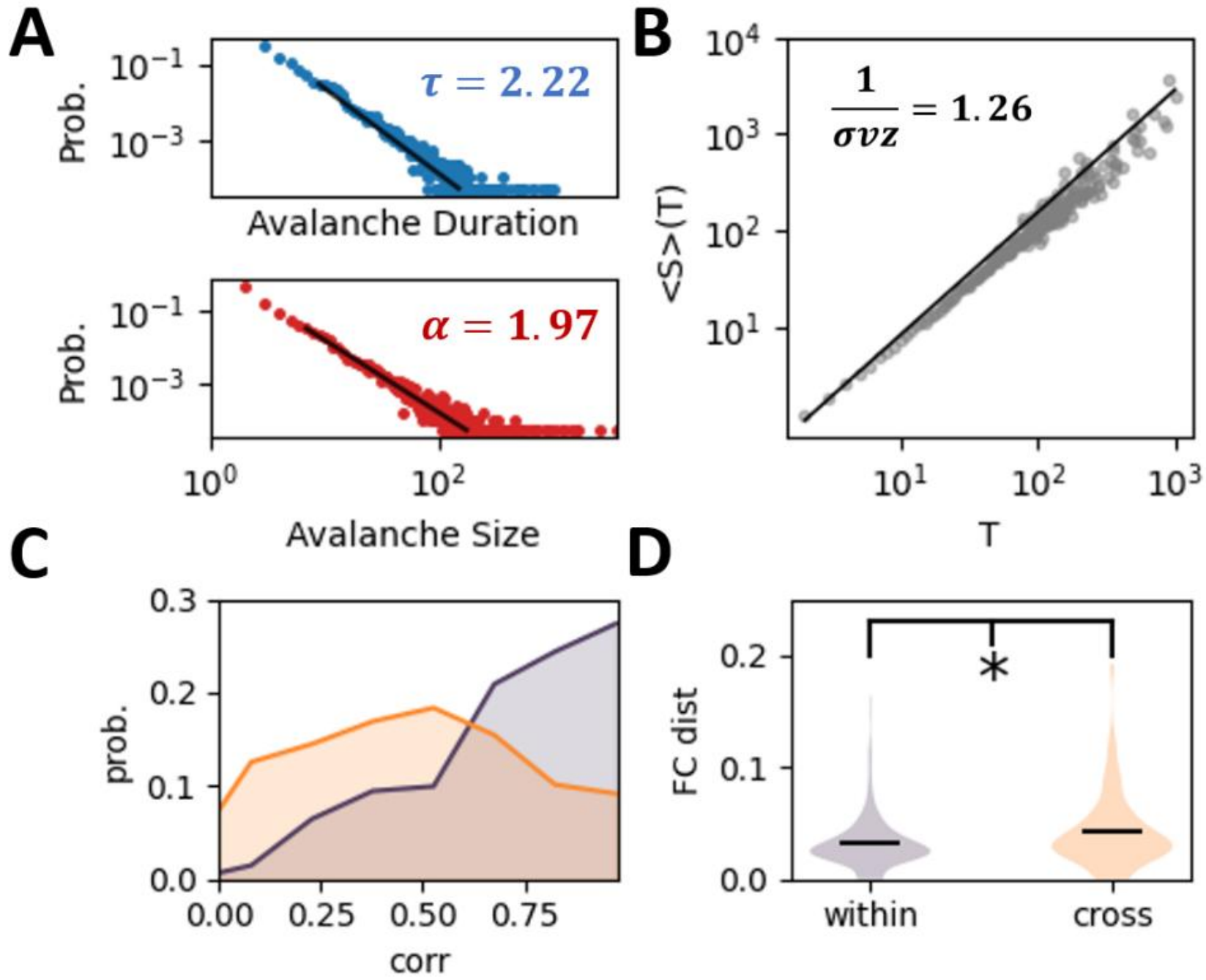

**Figure 1. Variable spontaneous dynamics in the experimental data. (A)** The power-law distributions of both avalanche duration and size of the experimental data in the spontaneous state. (**B**) Average avalanche size $\langle S \rangle$ given avalanche duration $T$. The relationship between the three power-law exponents aligns with theoretical predictions, i.e., $\mathrm{DCC} = |1/\sigma\nu z - (\tau - 1)/(\alpha - 1)| = 0.05$. The black line refers to the theoretical prediction with the slope equal to $(\tau - 1)/(\alpha - 1)$. **(C)** The probability distribution of correlation of mean firing rate of neurons within-session (orange curve) and cross-session (olive curve). (**D**) The distance between the functional connectivity, i.e., correlation between the neuronal activity, within-session and cross-session (t-test, $p = 0.025$, see Method for detail).

Criticality implies variable active states, demonstrated by a high susceptibility and entropy [32]. Here, we quantified the variability of activity patterns in the spontaneous state by evaluating the mean firing rate of neurons $v \in R^N$ ($N$ represents the number of neurons) across trials, and calculating the across-trial correlation of $v$ within and across sessions. In experimental data, we confirm smaller cross-session correlations, indicating changing spontaneous activity patterns (Fig. 1C). We further evaluate the noise correlation in the spontaneous state of neurons, which has been considered to reflect the underlying structural connectivity pattern. The across-trial distance of noise correlation matrices within session is smaller than that of the cross sessions (t-test, p =0.02, Fig. 1D), which indicates the fluctuation of the underlying effective neuronal connectivity. To highlight the changing network structure, we directly estimate the effective connectivity (see Method) from the neuronal activity, where the within-session effective connectivity is more similar than the cross-session one (Fig. S1).

The key question is whether there are reliable response patterns from such highly variable critical spontaneous states to support the reliable behavior, especially with changing network connections across sessions due to plasticity. To answer this question, we began with a detailed analysis of the experimental data previously used to study representational drift [8,36]. We first examined single-neuron representations, generally considered unstable, by checking the consistency of active neurons for a given stimulus, defined as those with above-mean firing rates. For visualization, we used one mouse as the example to illustrate the following results, and checked the statistical significance in all experimental data (Fig. S2). As expected, we found that nearly half of active neurons change their activation across sessions, indicated by non-overlapping regions between differently colored circles (Fig. 2A). We next examined reliability at the population activity, i.e., how the combination of all neuron responses to the given stimuli, where considerable drift has also been observed. We evaluate the population activity by the mean firing rate in response to the given stimulus across trials, named as mean population response vector [36]. Consistent with previous results, we found that the correlation between mean population response vectors across sessions in the experimental data is quite small (Fig. 2B), indicating drift in representation of fixed signals.

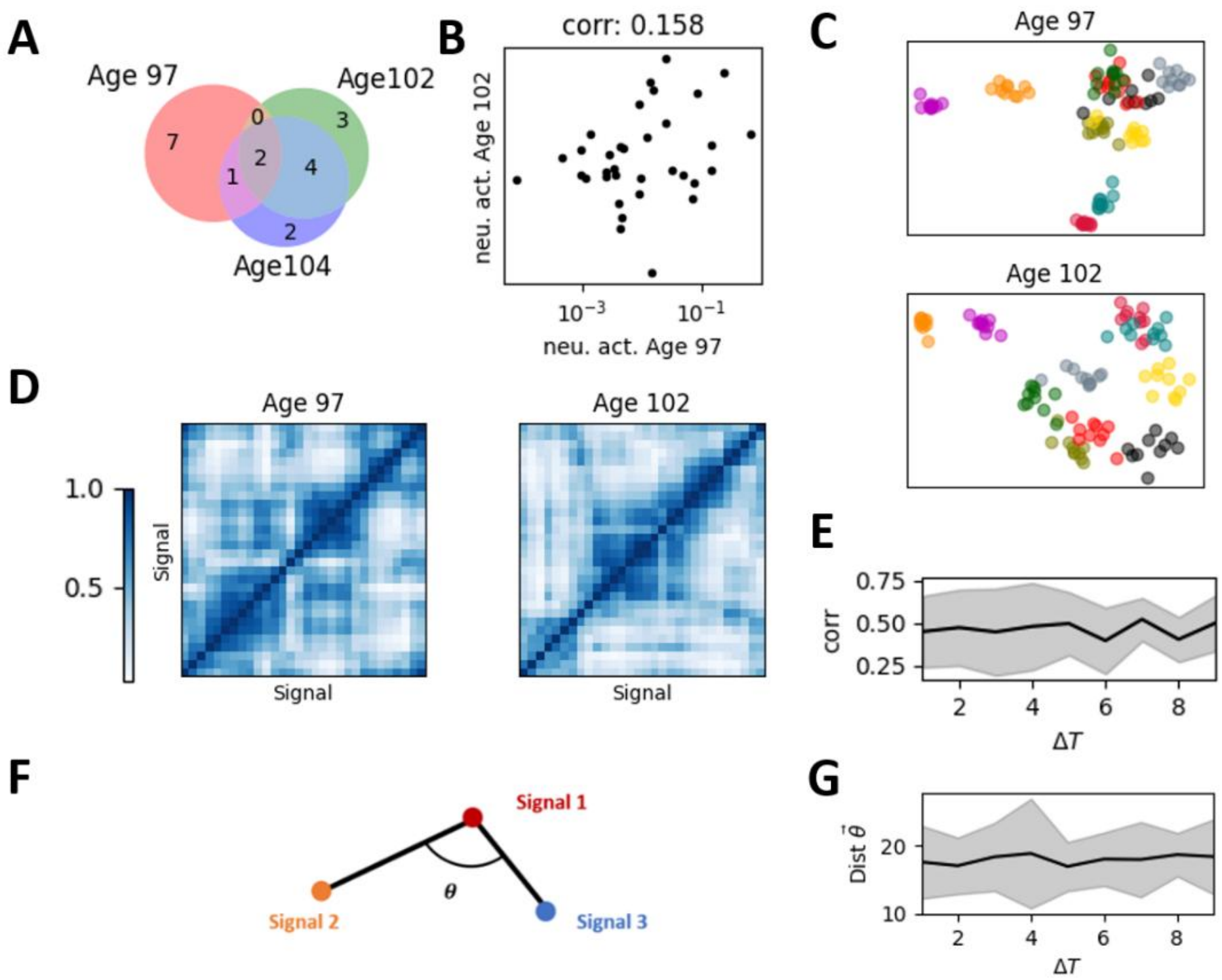

**Figure 2. Representational drift and the restricted representational geometry in the experimental data. (A)** The set of active neurons for a given stimulus changed across sessions (days). Red, green, and purple dots represent active neurons on the corresponding sessions, with overlapping colors indicating neurons active across multiple sessions. **(B)** The correlation between the mean population response vector across sessions of the same signal. **(C)** The low-dimensional representation of the population response vector of different signals and different trials in two sessions (top panel and bottom panel, respectively). Each color represents a different stimulus, with individual points of the same color representing different trials of that stimulus. **(D)** The representation similarity matrix, i.e., the correlation matrix between mean population response vectors for all stimulus pairs within session. **(E)** The cross-session correlation of the representation similarity matrices across signals obtained on different sessions (e.g., the two matrices shown in Figure 3D) as a function of the temporal distance between these sessions. **(F)** Illustration of the representational geometry of population response with different stimuli (nodes), defined by the angles between edges of the mean population response, where each edge is denoted as difference between mean population response vectors of two signals (see Method). For reliable representational geometry across sessions, the angles between all the edges are supposed to be preserved. **(G)** Representational geometry undergoes subtle changes over time but remains confined within a restricted space, manifested by time-invariant drift distances.

Even though signal response patterns of a single stimulus turns to be variable, the structure of neuronal representations across multiple signals is reported to be relatively consistent [8]. We reproduced previous results showing that low-dimensional embeddings of mean population response vectors for different signals remain similar across sessions (see Method). As shown in Fig. 2C, the neural representation of two signals that are closer (like the blue and green clusters) on one session (top panel) may tend to remain closer on subsequent sessions (bottom panel). We quantified the reliability of neuronal representation structure of multiple signals using representation similarity matrices, which measure correlations between mean population response vectors of different signals within sessions (Fig. 2D). For each session, we obtained a representation similarity matrix, then calculated the correlation (or distance) between these matrices as a function of the time difference $\Delta T$ between sessions. This correlation typically takes value about 0.4, indicating a drift of representation. We found that, across-sessions, the change of correlations of representation similarity matrices is nearly independent of $\Delta T$, consistent with findings in human visual cortex [11]. Intuitively, such drift can be considered as a rotation of low-dimensional representations as in Fig. 2C, where nodes (centers of the same-color clusters) change their positions across sessions (top panel and bottom panel), but their relative positions remain stable. To illustrate this overall rotation with stable internal structure and evaluate such relationship between the mean population response vectors of different signals within sessions, we characterize the geometry of the mean population response vectors using the angles between the edges linking mean population response vector pairs (see Method, and illustration in Fig. 2F) and found that representational geometry (distance of all the angles) remains restricted across sessions, as demonstrated in Fig. 2G.

In short, following established experimental measurements, we confirm that despite changes in both single-neuron and population response patterns over time, their representational similarity and representational geometry converge to a restricted space, as their changes are nearly invariant to the duration, which prevents the representations in the distant sessions becoming highly different.

These results provide direct evidence that representational drift and certain degree of reliability in the evoked states coexist with the critical spontaneous activity of the same circuit. Next, we investigated using modeling how these two pronounced aspects of neural dynamics with and without stimulations can be reconciled and whether the critical spontaneous activity plays nontrivial roles.

**Self-Organized Criticality and Representational Drift Emerge through Homeostatic Plasticity**

To study how these spontaneous dynamics may influence the signal representation in the changing networks (Fig. 3), we first needed to build a biologically realistic network model that captures essential dynamic features. As experimental data suggested changes of the underlying network connectivity across trials, we assumed plasticity is involved. Therefore, we used a biologically plausible homeostatic plasticity rule which has been explored in previous computational study [35,37], and developed an excitation-inhibition network model incorporating this plasticity rule at connections from inhibitory to excitatory neurons. This inhibitory spike timing-dependent plasticity (iSTDP, equation 2 in Methods) regulates the synaptic strength (Fig. 3A), and constrains the background noise-driven network from a random initialization of connectivity to a relatively stable configuration with small fluctuations in connectivity weights (Fig. 3A). The spontaneous dynamics self-organize into a critical state, characterized by power-law distributions of avalanche size and duration (Fig. 3C), with the relationship between these exponents predicted by the crackling noise theory (Fig. 3D). Similar to previous work [35], the self-organized criticality (SOC) mechanism in our model produces a critical line in the model parameter space where initial inhibitory synaptic strength does not influence criticality, while changing the inherent timescale of inhibitory synapses results in phase transitions, as we will discuss in the subsequent sections.

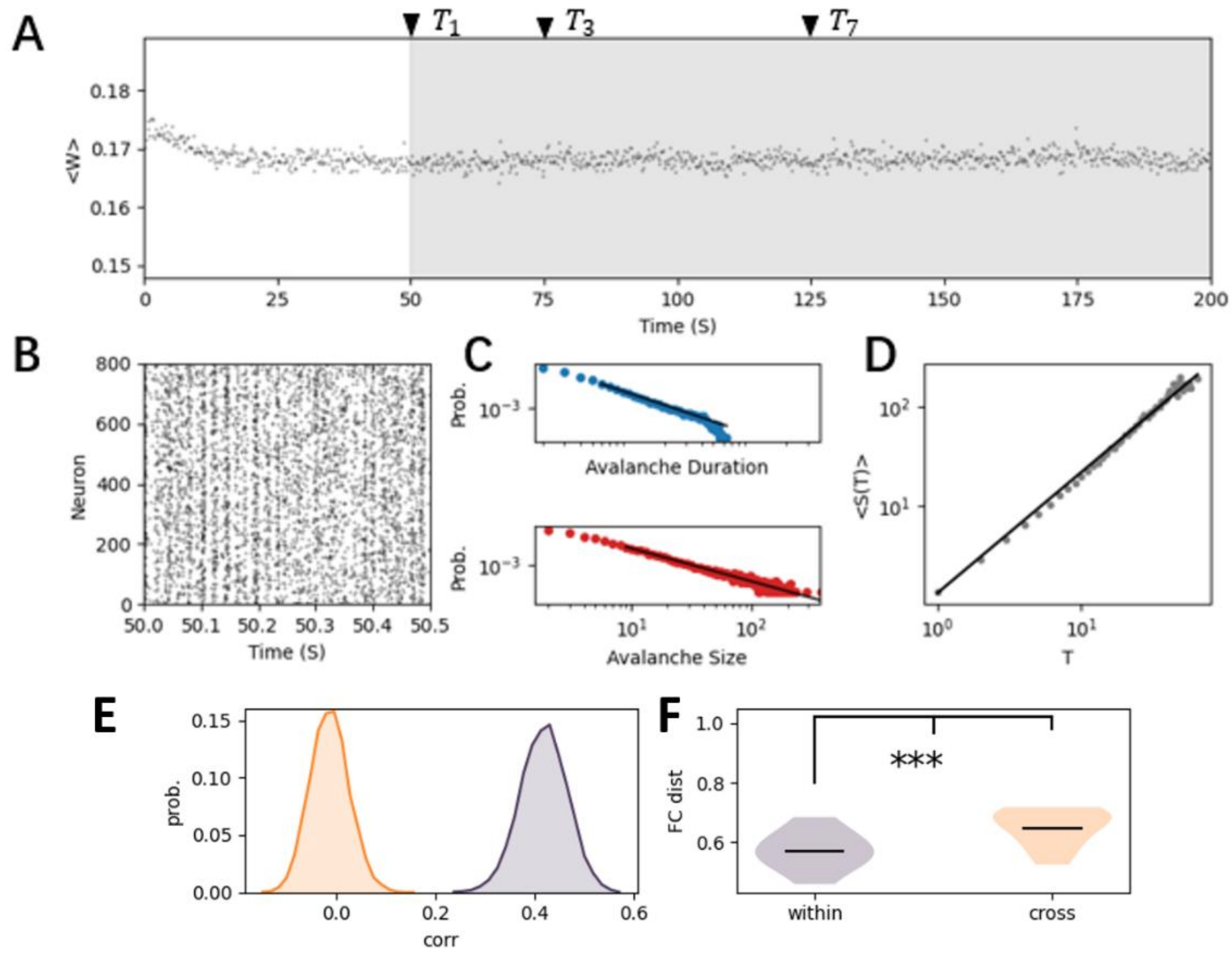

**Figure 3. The self-organized criticality (SOC) in the E-I network with homeostatic plasticity. (A)** The synaptic strength dynamics of a randomly initialized network with homeostatic plasticity, converged to a relatively stable state characterized by a nearly constant mean synapse strength from inhibitory neuron to excitatory neuron, as shown in the grey region. To study representational drift in this model, we selected network configurations at several time points ($\{T_i\}_{i=1,2..10}$) within this stable state. These time points are referred to sessions in experimental data. **(B)** The raster plot of neural activity in the stable state. **(C)** The near power-law distribution of the avalanche duration (blue) and avalanche size (red), indicating that the homeostatic plasticity organizes the neural dynamics into the (near) criticality. **(D)** The relationship between the average avalanche size with respect to avalanche duration follows the theoretical prediction ($\mathrm{DCC} = |1/\sigma\nu z - (\tau - 1)/(\alpha - 1)| = 0.04 \pm 0.016$). **(E)** The probability distribution of correlation of mean firing rate of neurons within (olive curve) and across (orange-curve) sessions. **(F)** The distance between the functional connectivity, i.e., correlation between the neural activity, within and cross-session.

Our plasticity model reproduces this variable critical spontaneous pattern, and the underlying fluctuations in network connectivity due to plasticity under critical spontaneous activity with background noise (Fig. 3 E, F, and Fig. S1). Networks with identical connectivity (mimicking same session in experimental data) naturally produce more similar activity patterns than those with changed connectivity patterns cross-session. Overall, we demonstrated that the variable spontaneous dynamic features in our SOC model are qualitatively consistent with experimental data.

We next study signal representation in our SOC model. We set the encoded signal as a 6 Hz frozen Poisson spike train for each neuron, with different signals corresponding to different stochastic samples of the Poisson signals (see Methods). We compared network configurations at different time points (T1, T2 in Fig. 3A) after convergence to SOC to simulate neuronal networks on different sessions. We then introduce different frozen Poisson signals into the networks of different sessions, each for 100 trials, and record population response vectors for each trial and signal at each session.

We performed the same analyses used for experimental data, with results shown in Fig. 4. Despite changing connectivity patterns induced by the interplay between the plasticity rule and spontaneous dynamics, these networks share common response dynamics properties similar to the experimental data: variable representations across sessions at both individual neuron and population response (Fig. 4A-C), and the restricted representational geometry (Fig. 4D-E).

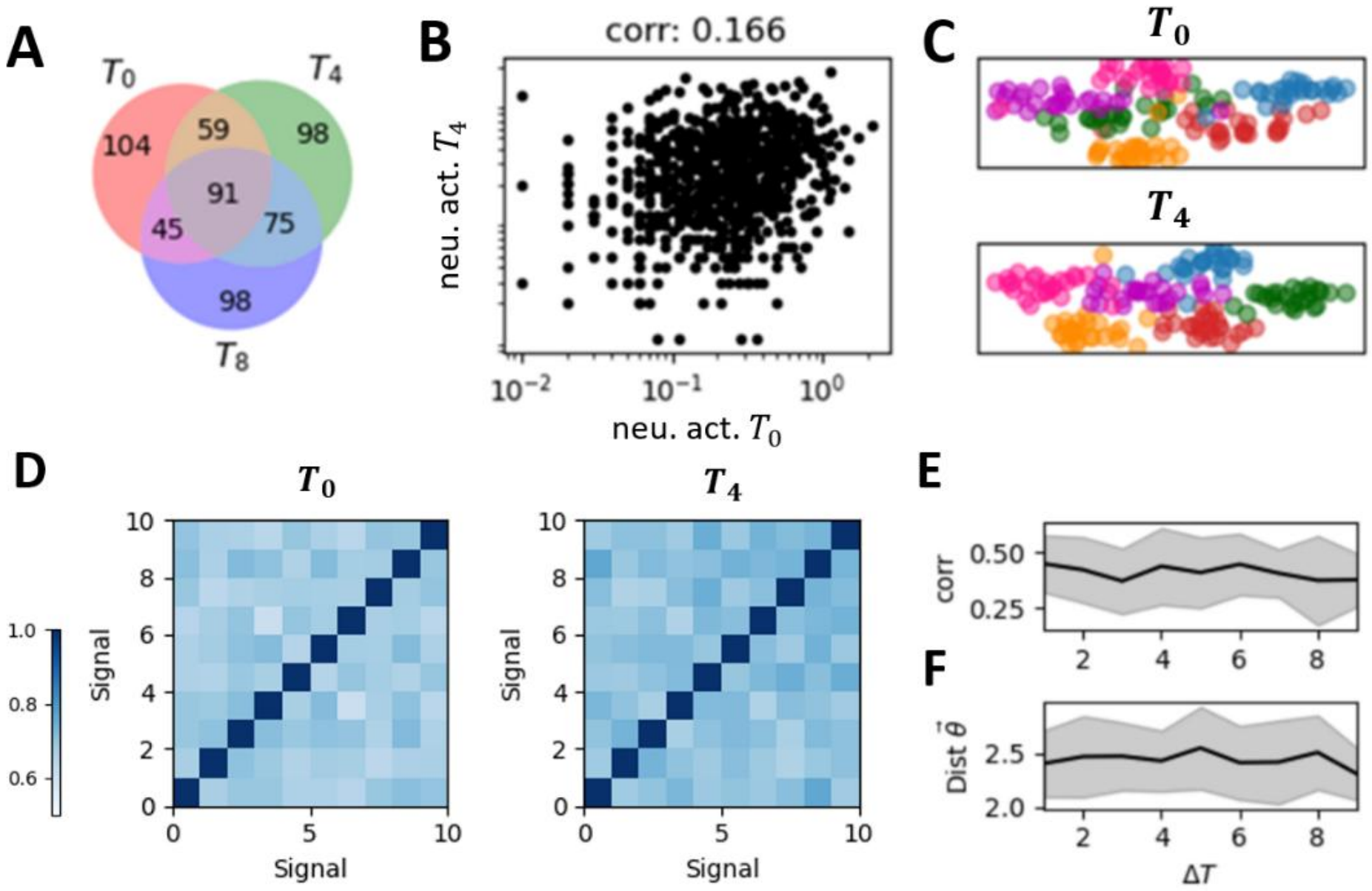

**Figure 4. Representational drift and the restricted representational geometry in the self-organized critical (SOC) model. (A)** The set of active neurons for a given stimulus changed across sessions. Red, green, and purple dots represent active neurons on the corresponding sessions, with overlapping colors indicating neurons active across multiple sessions. **(B)** The correlation between the mean population response vector across sessions of the same signal. **(C)** The low-dimensional representation of the population response vector of different signals and different trials in two sessions (top panel and bottom panel, respectively). Each color represents a different stimulus, with individual points of the same color representing different trials of that stimulus. **(D)** The representation similarity matrix, i.e., the correlation matrix between mean population response vectors for all stimulus pairs. **(E)** The cross-session correlation of the representation similarity matrices across signals obtained on different sessions (e.g., the two matrices shown in Figure 4D) as a function of the temporal distance between these sessions. **(F)** Representational geometry undergoes subtle changes over time but remains confined within a restricted space, manifested by time-invariant drift distances.

Although plasticity is driven by the spontaneous dynamics, the resulting changes of synapse strength are not random. Since the homeostatic plasticity rule regulates excitatory neuron's firing rates to given values, connection strengths from inhibitory neurons to highly active neurons are more likely to be enhanced rather than reduced, which constrains the freedom of synapse strength $W$. To examine how homeostatic plasticity differs from random plasticity in terms of signal representation, we obtained the distribution of $\Delta W_{ij}$ from numerical simulations of self-organized networks after dynamics convergence to SOC (gray region in Fig. 3A), as shown in Fig. 5A. We then built a network model with randomly shuffled plasticity where the individual synaptic strength change at each time step is randomly sampled from the $P(\Delta W_{ij})$ distribution rather than being determined by neural spike timing. Such random plasticity shifts the network to a slightly subcritical state from the original critical state, where neural avalanche size deviates from power-law distribution (Fig. 5B). We found that the representational drift accumulates with sessions, the representational similarity reduces continuously and the representational geometry is no longer restricted (Fig. 5C). In summary, we showed that SOC induced by homeostatic plasticity can maintain reliable representations of external signals despite representational drifts.

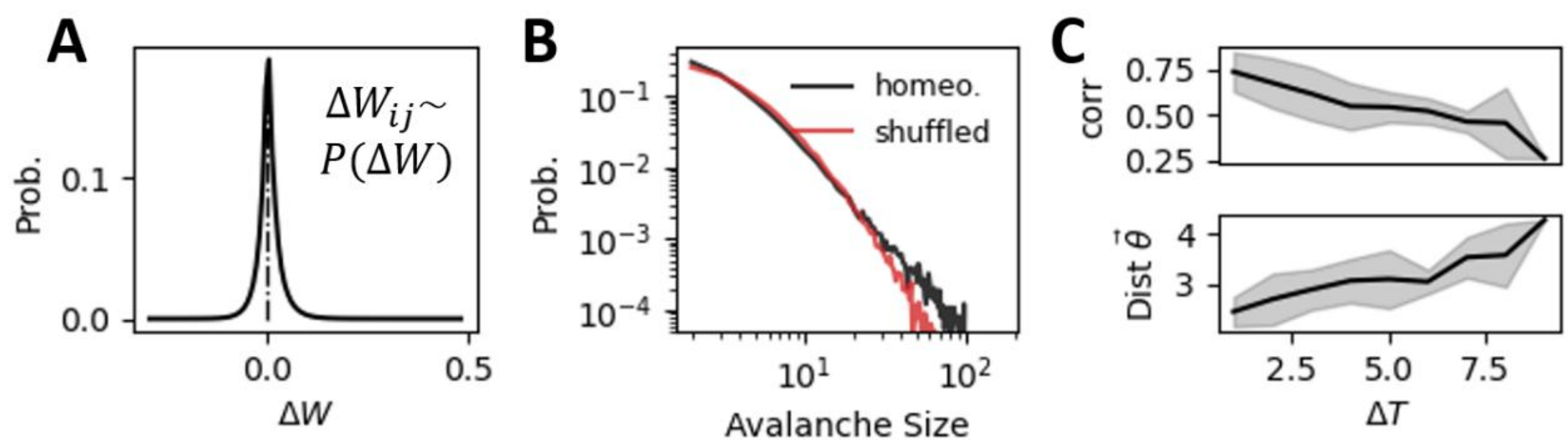

**Figure 5. Random shuffled plasticity biases the network from criticality and breaks down restricted representational geometry.** **(A)** The distribution of synaptic weight changes ($\Delta W_{ij}$) obtained from numerical simulations of the self-organized critical network with homeostatic plasticity (as described in Figure 2A). To create the "random shuffled plasticity" condition, we replaced the original homeostatic plasticity rule with random sampling from the empirically observed $\Delta W_{ij}$ distribution. Each connection was updated independently using this distribution. **(B)** The shuffled plasticity shifts the network into a slightly subcritical regime compared to the near-critical dynamics observed with homeostatic plasticity. This shift is further supported by an increase in DCC value for the shuffled plasticity network compared to the self-organized critical network with homeostatic plasticity ($\mathrm{DCC}_{\mathrm{homeo}} = 0.040 \pm 0.016, \mathrm{DCC}_{\mathrm{shuffled}} = 0.072 \pm 0.022$). **(C)** The randomly shuffled plasticity disrupts the restricted representation geometry observed with homeostatic plasticity. This is evidenced by a reduction in representational similarity across time (upper panel) and an increase in the distance between representational geometries (lower panel).

### Enhanced Neural Representational through Self-Organized Criticality and its Network Mechanism

Self-organized criticality implies that the network model evolves into the critical state independent of the initial inhibitory strength distributions. However, the dynamics of the model and the phase transition can still be determined by several control parameters. Here, we set the timescale of inhibitory synapses $\tau_d^I$ as the control parameter, following previous studies [33,38] . For a given $\tau_d^I$, starting with an arbitrary initial inhibitory connectivity strength distribution, the network with homeostatic plasticity will evolve to its stable dynamic state. We demonstrated that increasing parameter $\tau_d^I$ induces a clear phase transition, as shown in Fig. 6A and Fig. S3, from an asynchronous subcritical state (Fig. S3, left panel) to a synchronous supercritical state (Fig. S3, right panel), through a critical state characterized by intermediate level of sparse synchrony (Fig. S3, middle panel) and its avalanche duration distribution can be best fitted by a power-law distribution (Fig. 6B).

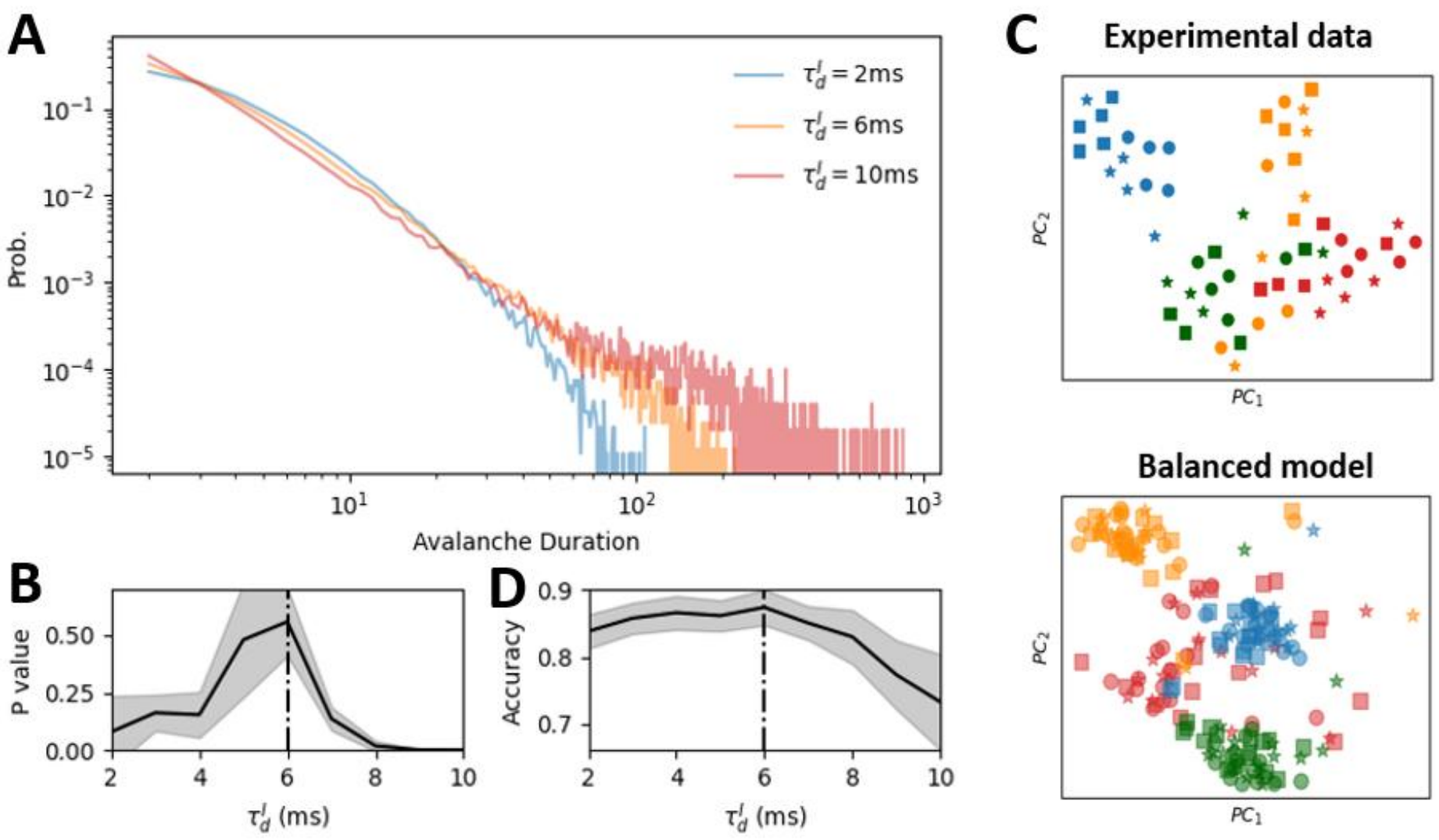

**Figure 6. Self-organized critical (SOC) state maximizes cross-session low-dimensional representation reliability. (A)** The distribution of avalanche durations for three representative values of $\tau_d^I$. **(B)** The p-value obtained from comparing each avalanche duration distribution to the optimal power-law distribution. A higher p-value indicates a closer fit to the power law, suggesting closer proximity to criticality. The network with $\tau_d^I = 6\,ms$ (black dashed line) exhibits the closest fit to the power law, indicating a near-critical state. **(C)** Low dimensional representation of the linear discriminant analysis (LDA) results for four example stimuli (represented by different colors). Circles, stars and squares denote the trails of three different sessions. The close alignment of markers for the same stimulus cross-session indicates that the LDA decoder successfully captured stimulus-relevant features that are invariant to session-to-session variability. The upper and low panels show the results for the experimental data and model simulation, respectively. **(D)** The performance of the LDA decoder as a function of $\tau_d^I$ in the model. Decoding performance was quantified as the classification accuracy. The SOC state ($\tau_d^I$ = 6 ms) exhibited the highest cross-session decoding performance. The shade area in (B) and (D) represents the standard deviation across 10 network realizations.

The mechanism underlying this phase transition can be understood as follows: increased $\tau_d^I$ prevents inhibitory currents from promptly canceling excitatory currents, creating a time window for neurons to fire together and having a higher chance to generate a more strongly synchronous activity under a given excitability. Such co-spiking between neurons later produces substantial increases in inhibition strength, which then inhibits a large population of neurons below their target firing rates and network becomes silent for a short duration (Fig. S4). Due to background input, the neurons gradually spike and reduce the inhibition strength. Once the excitability of network returns near the target value, then the neurons will have larger chance to co-spike. This co-spike will again cause a cyclical activity pattern. When $\tau_d^I$ is small, the inhibitory current can quickly counteract the excitatory current with few neurons, thereby maintaining the network's excitability close to the target value without inducing network-level collective activity. Only when $\tau_d^I$ is moderate (e.g., 6 ms), the network converges to the target excitability, exhibiting sparse synchrony near the transition from asynchrony to full synchrony

To directly evaluate neural representation across sessions, we use linear discriminant analysis (LDA) (see Methods) to train a common decoder to identify each signal by using response activities of the same signal from different sessions. Generally, LDA returns a discrimination vector for each signal; we added another biologically reasonable constraint to make these discrimination vectors nonnegative. By projecting activity onto the discrimination vectors, we showed that the common decoder can captures activity features of the same signal across networks of different sessions in the experimental data (Fig. 6C, the decoding accuracy for all subjects in experimental data is $0.61 \pm 0.15$). In Fig. 6C, points with the same color are results for the trials of the same signal and different shapes represent different sessions. Importantly, the decoding accuracy in the model peaks near the critical state (Fig. 6D). We further examined the results when the input stimuli have different durations, and also tried different decoding schemes, including common decoder and transfer decoder (see Methods). The critical state robustly outperforms subcritical and supercritical states (Fig. S5) in all cases. This result indicates that the criticality enhances the cross-session low-dimensional representation reliability.

We next investigated the evolution of inhibitory synapse by examining changes in synapse strength patterns and indegree patterns during the self-organizing process in the presence of the homeostatic plasticity. The synapse strength patterns refer to the average strength of all inhibitory synapses, while the indegree pattern refers to the average total inhibitory synapse weights of each excitatory neuron. Both changes of synapse strength (evaluated by the L1 distance between different sessions, see Methods) and indegree patterns converge to static non-zero values after a transient period (Fig. 7A), suggesting that homeostatic plasticity-induced connectivity structures show long-term stability, where the difference of connection between sessions separated by large time interval is similar to that by medium time interval, which shows the restricted representational geometry. In contrast, the randomly shuffle plasticity leads to the continuously increasing differences for both synapse strength and indegree pattern (Fig. S6). We next examined how phase transitions governed by $\tau_d^I$ influence this restricted structural similarity. The changes of synapse strength and indegree at the stable state after the transient are both minimized near the critical state (Fig. 7B), which thus has enhanced reliable representation. Overall, we showed that SOC contributes to reliable representations in changing networks by restricting the drift of connection structures and response dynamics.

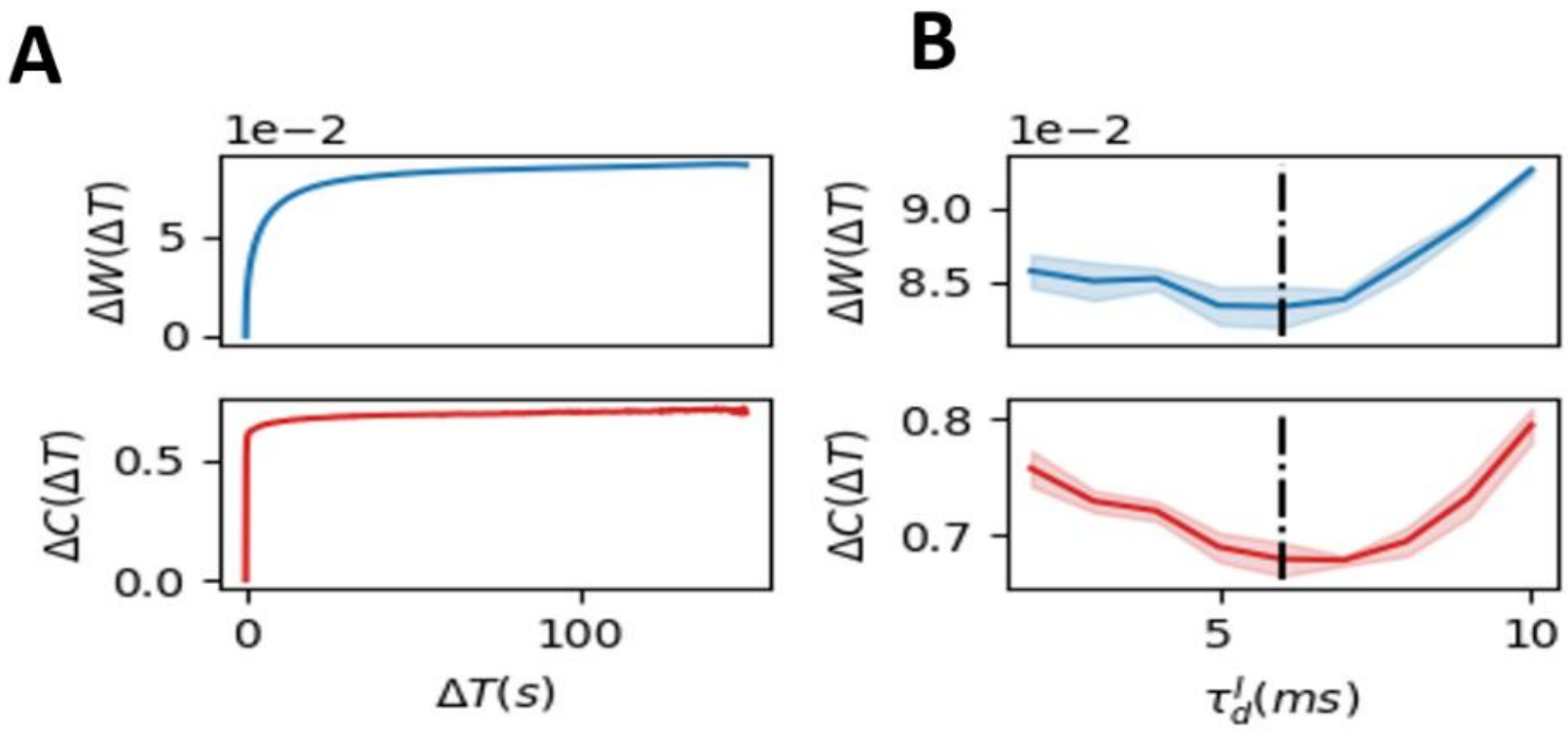

**Figure 7. Structural mechanism underlying restricted representation.** (**A**) The change in synaptic weight patterns ($\Delta W$, blue) and the change in inhibitory degree ($\Delta C$, red) for each excitatory neuron over time in the SOC network ($\tau_d^I = 6\ ms$). Both $\Delta W$ and $\Delta C$ rapidly stabilize after a brief period of initial fluctuations. (**B**) The static state value of $\Delta W$ and $\Delta C$ as the functions of $\tau_d^I$. The near-critical state ($\tau_d^I = 6\ ms$) exhibited the lowest $\Delta W$ and $\Delta C$ indicating that the network structure is most stable at criticality.

## Discussion

A key question in understanding neural representations across a long timescale is how neural networks maintain reliable response patterns under the inevitably plasticity induced by the self-organized and highly variant spontaneous dynamics. Here, we propose a self-organized criticality neural network model capable of reproducing both spontaneous critical states and representational drift with restricted representational geometry manifesting representational reliability, as observed experimentally. Our analysis suggests that the critical state constrains the fluctuations of network connectivity within a narrow regime, thereby facilitating reliable low-dimensional representation despite ongoing plasticity.

In our study, SOC in an excitatory-inhibitory (E-I) network is achieved through biologically plausible homeostatic plasticity of inhibitory connections [22,35]. The concept of inducing SOC via homeostatic plasticity has previously been explored in various model frameworks [23,24], as homeostatic plasticity inherently adjusts network excitability towards a stable value which can be put near the critical bifurcation point. In our implementation, inhibitory spike-timing-dependent plasticity (iSTDP) is employed to realize homeostatic plasticity, similar to previous computational studies, by modulating network excitability through inhibitory synapse [35]. SOC emerges in our model when the decay timescale of inhibitory receptors is moderate, controlling the co-activation of excitatory neurons [39]. The bifurcation observed here likely corresponds to a transition from asynchronous (Fig. S3, blue) to synchronous (Fig. S3, red) network states. In a simplified E-I model without plasticity, this control parameter has been shown to induce a Hopf bifurcation, where irregular spikes of individual neurons coexists with the scale-free avalanches [33,34,38].

Representational drift, which challenges the traditional assumption of stable neuronal tuning over prolonged periods, has been reported across multiple brain regions, species, recording techniques, and spatial scales [6,7]. Drift in individual neuronal responses and population response vectors has been commonly observed in sensory cortices, including the visual [8] and olfactory cortices [9], higher-order regions such as the posterior parietal cortex [5], and subcortical structures like the hippocampus [40] in mice, and also including the primary visual cortex in humans [11]. While representational similarity and geometry of multiple signals also exhibit drift in the olfactory cortex [9], some studies indicate a relatively restricted geometry in the visual cortex and hippocampus [8,11,40]. This observation aligns with our analysis of experimental data from primary visual cortex (V1) during passive visual information representation. The inconsistency in the drift geometry across different brain regions remains poorly understood and may reflect inherent differences in regional stability or the increasing complexity of signal representation along the cortical hierarchy. A recent study using simultaneous multi-region recordings revealed heterogeneous stability across cortical areas during go/no-go tasks, with visual cortex representations remaining relatively reliable (as evidenced by higher accuracy of common decoders), whereas motor cortex representations exhibited relative instability [10]. Task complexity also influences representational stability: simple stimuli, such as drifting gratings, elicit reliable representations across sessions, whereas complex stimuli, such as natural movies, induce representational drift [41]. As the higher order brain regions in the cortical hierarchy carries more complex cognitive function [42,43], they could also contribute to the instability across sessions.

Our SOC model successfully reproduces the commonly observed representational drift while capturing the restricted representational geometry observed experimentally in visual cortex data. Combined with findings that random plasticity alone leads to purely representational drift and that optimal cross-session reliable low-dimensional representation occurs near criticality, our results suggest that the representational stability exhibiting heterogeneity across brain regions may reflect that the spontaneous states in different brain regions have different proximity to criticality [44,45].

How reliable task-related information emerges from variable neural representations? Two hypotheses have been proposed: the coding null-space hypothesis and the co-varying decoding hypothesis [6]. The coding null-space hypothesis suggests that representational drift occurs exclusively within task-irrelevant subspaces, which requires a supervised learning mechanism precisely counteracts random fluctuations to maintain orthogonality between drift and representation spaces. This strong assumption has been challenged by experimental evidence indicating that decoders trained across multiple sessions perform worse than session-specific decoders, implying drift occurs within task-relevant representational spaces as well [46]. Nevertheless, representational spaces indeed exhibit less drift compared to noise-variance spaces, highlighting their relative stability [46]. This relatively stable representational space could emerge from biologically plausible learning rules; for example, computational models employing Hebbian and anti-Hebbian rules to optimize representation similarity successfully reproduce representational drift [12], and recent study implementing the STDP for fast learning to reduce the representational drift [47]. A more intuitive analogy is training simple deep neural networks for image classification [13,36]: although the loss function converges to near-zero values, stochastic gradients yield multiple networks with similar performance but variable neuronal tuning curves, demonstrating representational drift alongside stable task performance. Unlike these models that mainly constrain task states without explicitly considering the contribution from spontaneous states, our SOC model regulates mean firing rate in the internal spontaneous state which constrains representational geometry in the task state. The mechanism is that the underlying network connection is driven into a regime avoiding the further accumulation of the drift and the changes in network connectivity strength in this regime is smallest at the SOC state. Instead of considering the perfectly stable representation space, we propose that a restricted representation space observed experimentally can be easily obtained by simple unsupervised plasticity rule that underlies the co-evolution of network connectivity and spontaneous activity. Alternatively, the co-varying decoding hypothesis posits that despite changing representations, stable task performance can be maintained if downstream decoder networks receiving feedforward inputs adapt their synaptic connections via appropriate plasticity rules (e.g., Hebbian [46] or spike-timing-dependent plasticity [48]). In this context, our SOC model exhibits relatively stable yet flexible response pattern comparing to the noncritical states, potentially facilitating the rewiring in downstream decoder networks for a stale decoding.

Critical systems typically exhibit highly variable and sensitive dynamics [32], seemingly contradicting the requirement for stable representations. In the classical branching model, the critical state is characterized by maximal dynamic range and sensitivity to small perturbations [26,27] , as confirmed in in vivo experiments [29,30] . As expected, this simplest non-equilibrium model also exhibits minimal reliability to fixed input signals. Reliable critical states have been identified in E-I balanced networks, demonstrating stimulus-induced reductions in trial-to-trial variability [34,49] and near-optimal low-dimensional representations [33]. Here, we extended this framework by incorporating homeostatic plasticity to achieve a reliable critical state, revealing constrained representational geometry at criticality. By showing that more variable synapse dynamics happen in either random shuffled plasticity or in the self-organized non-critical states, we highlight that the critical spontaneous dynamics in the E-I network provide an effective and reliable low-dimension representation for the downstream neurons to decode the stimulus against both the random background noise and the variation in the network connection due to the homeostatic plasticity.

Our model yields several experimentally testable predictions. First, the low-dimension critical representations can be directly validated by perturbing the cortex to drive it away from the critical regime, and measuring resultant changes in cross-session representation [30]. This perturbation can be done either pharmacologically [30], or through the visual depravation [23]. Second, advances in multi-region recording [6] allows further validation on whether heterogeneous representational stability across brain regions correlate with their locally distinct spontaneous dynamics [44,50]. In summary, integrating empirical data analysis and computational modeling, we demonstrated that representational drift and restricted representational geometry is modulated by spontaneous dynamics, and the apparently most variable critical state can counter-intuitively facilitate long-term reliability low-dimension representation. This framework bridges neural representation during the evoked state with spontaneous activity in the absence of stimulations, offering insights into how the baseline states of the brain may affect cognitive function and brain disorders. Future work may further study how the representational drift contributes to cognitive flexibility while maintaining performance stability, where critical spontaneous state may be highly relevant for its variability and sensitivity to perturbations in a constantly changing environment.

## Methods

### Experimental data

We utilized data from the Allen Institute's Visual Coding 2-Photon dataset (https://portal.brain-map.org/circuits-behavior/visual-coding-2p). This comprehensive dataset contains both spontaneous and stimulus-evoked neural activity recorded from mouse visual cortex across multiple sessions using two-photon calcium imaging [8]. Our study involved data from 78 mice, with recordings spanning three days (sessions). For the spontaneous activity, mice were presented a fixed gray screen ( $0.248 \pm 0.002$ hours for each mouse). For evoked activity, mice were presented with repeated 30-second natural movie stimuli. The recorded neuron comes from the primary visual cortex across all layers, as descripted in [36]. To study spontaneous neural dynamics and due to short recording interval of the spontaneous state, we aggregated spontaneous activity recordings from all mice across all experimental sessions to perform avalanche analysis, providing a robust dataset for examining intrinsic network properties. To investigate external signal representation, we followed established preprocessing methods from previous studies [8,36]. Specifically, we divided each 30-second natural movie into 30 distinct clips, treating each 1-second clip as a unique signal type. For each clip, we calculated the population response vector during the corresponding time window, which served as the neural activity representation for that particular signal. When we evaluated the relationship between duration of days and representational similarity and representational geometry, we removed the cases with duration larger than 10, as there are fewer than three case for each large duration of days.

### Spiking neural model

We consider an excitation–inhibition neural network with conductance-based leaky integrate-and-fire neural dynamics. This model can unify several biologically observations of neural dynamics across different scales, including the irregularity of individual neuronal spikes, weak synchrony of neuronal spiking, collective oscillations at the population level and scale-free avalanches, and it has also been used to study the neural response reliability [33,34]. Our model comprises $N$ = 1000 randomly connected neurons with connection probability $p = 0.2$, where 80% of neurons are excitatory and 20% are inhibitory. The dynamics of system is determined by the membrane potential $V_i$ of each neuron, where

$$\frac{dV_i}{dt} = \frac{1}{\tau_\alpha}(V_{\text{rest}} - V_i) + I_i^{\text{rec}} + I_i, \alpha \in \{E, I\}. \tag{1}$$

In the absence of recurrent input $I_i^{\text{rec}}$ and external input $I_i$, the linear leak term will drive the dynamics into a resting potential at $V_{\text{rest}}$. When the neuron $i$ receives enough drive and $V_i$ growths beyond a given threshold $V_{\text{th}}$, a spike is then generated and transmitted to its output neighbors. We then reset the membrane potential to $V_{\text{reset}}$ and the neuron enters into a refractory period of 2 ms (excitatory neuron) or 1 ms (inhibitory neuron). We described the spike activity of neuron $i$ as the spike train $S_i = \sum_k \delta\left(t - t_i^k\right)$, where $t_i^k$ is the $k$-th spike time for neuron $i$. In the conductance-based model, the recurrent input a neuron receives is modulated by the its own membrane potential, $V_i$, and is expressed as

$$I_i^{\text{rec}} = \sum_{\beta \in \{E,I\}} \left(V_\beta^{\text{rev}} - V_i\right) g_{\alpha\beta} F_\beta * \sum_{j \in C_i^\beta} S_j(t). \tag{2}$$

Here, $V_\alpha^{\text{rev}}$ is the reversal potential of the $\alpha$ (E or I) neuron and $g_{\alpha\beta}$ is the synaptic strength from the $\beta$ neuron type to the $\alpha$ neuron type, $C_i^\beta$ refers to the input neighbor of neuron $i$, and $F_\beta$ is the bi-exponential synaptic filter. $F_\beta$ is expressed as

$$F_\beta = \frac{1}{\tau_\beta^d - \tau_r}\left(e^{-t/\tau_\beta^d} - e^{-t/\tau_r}\right), \tag{3}$$

where $\tau_r$ and $\tau_\beta^d$ are the synaptic rise time and decay time for $\beta$ neurons, respectively. We set $\tau_r = 0.5$ ms and $\tau_E^d = 2$ ms to reproduce the fast excitatory synaptic dynamics generated by the AMPA receptor. The $\tau_I^d$ values are used as the control parameters for the critical transition, corresponding to the timescale of the $\text{GABA}_\text{A}$ receptor. $I_i$ in Eq. 1 represents the background input from other circuits. Each neuron in the spontaneous state receives the $N_o = p \cdot N_E$ independent Poisson spike train, $S^{\text{Poisson}}$, at a frequency $Q$, as background and we labeled the spike train of background input targeting on neuron $i$ as $\xi_i(t) = \sum_{k=1}^{N_o} S_{k,i}^{\text{Poisson}}(t)$; then, $I_i$ is given by

$$I_i = (V_E^{rev} - V_i) g_{\alpha o} F_\alpha * \xi_i(t). \tag{4}$$

**Homeostatic plasticity**

The homeostatic plasticity is implemented on the synapse from inhibitory neurons to excitatory neurons, $g_{EI}^{ij} \equiv W_{ij}$, where $j$ represents the source inhibition neuron and $i$ represents the target excitation neuron. We use the spike-timing-dependent plasticity rule, following study [35], where the synaptic strength $W_{ij}$ updates to $W_{ij} + \Delta W_{ij}$ as:

$$\begin{cases} \Delta W_{ij} = \eta(x_i - r_0) \ \text{ for presynaptic inhibition neuron spikes at time } t_j^f \\ \quad \Delta W_{ij} = \eta x_j \ \text{ for postsynaptic excitation neuron spikes at time } t_i^f \end{cases}. \tag{5}$$

In this equation, $\eta$ represents the learning rate, $r_0$ represents the depression factor and $x_i$ represents the synaptic trace as follows

$$\frac{dx_i}{dt} = -\frac{x_i}{\tau_{\text{STDP}}} + \Sigma_f \delta\left(t_i^f\right). \tag{6}$$

This plasticity rule can be viewed as a mechanism to constrain the activity of excitatory neuron near the depression factor $r_0$: if $(x_i - r_0) > 0$ and inhibition neuron spikes, then the inhibition connection will be enhanced, and verse vice.

To investigate the dynamics of connectivity structure driven by the spontaneous activity, we calculated the change of synapse strength as $\Delta W(\tau) = \langle \left\| \text{vec}\left(W(t+\tau)\right) - \text{vec}\left(W(t)\right) \right\| \rangle_t$, here $\|.\|$ represent the L1 distance, and we considered the change of indegree pattern as $\Delta C(\tau) = \langle \|C(t+\tau) - C(t)\| \rangle_t$, where $C_i(t) = \sum_j W_{ij}(t)$ is the total synaptic strength of incoming inhibitory neurons of the excitatory neuron $i$.

To study the representation of external stimulus in our model, we defined the signal as the frozen Poisson spike train into the network. On top of the background input, each neuron is given by an additional spike train $\xi_i^o(t) = \sum_{k=1}^{N_o} S_{k,i}^{\text{o}}(t)$, and $S_{k,i}^{\text{o}}$ is the *k* th Poisson spike train on neuron $i$ with the frequency $Q_o$. The stimulus signal lasts for 20 ms in the main text, which aligns with the characterized timescale of our model. We also change the duration of signal to examine the robustness of our results (Fig. S5).

The parameters we used in the model simulation are listed in the following table. The relative strength of excitatory connection and inhibitory connection follows the previous studies [38,51]. The parameters about the plasticity is from previous study [35].

| Parameter | Description | Value |
|---|---|---|
| $N$ | Total number of neurons | 1000 |
| $N_E$ | Excitatory neurons | $0.8 \cdot N$ |
| $N_I$ | Inhibitory neurons | $0.2 \cdot N$ |
| $p$ | Connection probability | 0.2 |
| $\mathrm{V_{rest}}$ | Resting membrane potential | -70 mV |
| $\mathrm{V_{th}}$ | Spike threshold | -50 mV |
| $\mathrm{V_{reset}}$ | Reset potential after spike | -60 mV |
| $\tau_E$ | Time constant for excitatory neurons | 20 ms |
| $\tau_I$ | Time constant for inhibitory neurons | 10 ms |
| $\mathrm{V_E^{rev}}$ | Excitatory reversal potential | 0 mV |
| $\mathrm{V_I^{rev}}$ | Inhibitory reversal potential | -70 mV |
| $\tau_r$ | Synaptic rise time | 0.5 ms |
| $\tau_E^d$ | Excitatory synaptic decay time | 2 ms |
| $\tau_I^d$ | Inhibitory synaptic decay time | 2-10 ms |
| $g_{EE}$ | E→E synaptic strength | 0.012 |
| $g_{IE}$ | E→I synaptic strength | 0.024 |
| $g_{II}$ | I→I synaptic strength | 0.31 |
| $g_{Eo}$ | External → E synaptic strength | 0.022 |
| $g_{Io}$ | External → I synaptic strength | 0.04 |
| $Q$ | Background Poisson frequency | 10 Hz |
| $Q_o$ | Poisson frequency of stimulus signal | 6 Hz |
| $N_o$ | Number of background inputs per neuron | $p \cdot N_E$ |
| $\eta$ | Plasticity learning rate | 0.02 |
| $r_0$ | Depression factor | 0.6 |
| $\tau_{STDP}$ | Synaptic trace time constant | 20 ms |

**Measurement of avalanches and criticality**

We evaluated neural avalanche properties in both experimental data and computational models. Since the experimental dataset consists of calcium signal time series rather than precise spike timing, we assessed avalanches from the mean time series $x(t)$ following the methodology described in [18]. Specifically, we defined a threshold as $T = \langle x \rangle + a \cdot \sigma^2(x)$, where $\langle x \rangle$ represents the mean signal value, $\sigma^2(x)$ represents the variance of the signal, and $a = 1$ is a coefficient. We also scan the parameter $a$ from $[0.5, 2]$, and find that our conclusions are robust. A neural avalanche was defined as a continuous period of activity exceeding this threshold. Avalanche duration *T* was measured as the length of this period, while avalanche size *S* was calculated as the summed above-threshold neural activity within this period. Using the Python package *powerlaw* [52], we estimated the optimal truncated power-law relationships: $P(T) \sim T^{-\alpha}$ and $P(S) \sim S^{-\tau}$ . According to crackling noise theory [53], the relationship between mean avalanche size $\langle S \rangle$ and for a given duration $T$ follows $\langle S\rangle(T) \sim T\hat{}(1/\sigma\nu z)$, where $1/\sigma\nu z = (\tau-1)/(\alpha-1)$ for critical state. We numerically fitted the exponent $1/\sigma\nu z$ from the avalanche data and then calculated the derivation coefficient from criticality [23] as $\mathrm{DCC} = |1/\sigma\nu z - (\tau-1)/(\alpha-1)|$. In agreement with previous study [23,54], we called a system is critical or near critical, if the scale-free avalanche, where the avalanches follow the three relations $P(T) \sim T^{-\alpha}$, $P(S) \sim S^{-\tau}$, $\langle S\rangle(T) \sim T\hat{}(1/\sigma\nu z)$, and $\mathrm{DCC} \sim 0$ is found in this system.

In our numerical simulations of spiking networks, we evaluated spike-based neural avalanches by first aggregating spike trains from all excitatory neurons, then converting this combined record into a vector where each element represented the spike count within a time window. The window size was set to $\mathrm{bin} = 1/\langle \mathrm{ISI} \rangle$, where $\mathrm{ISI}$ denotes the average inter-spike interval of the aggregated spike trains of all the neurons, following previous studies [16,38]. Within this vector, a neural avalanche was defined as a sequence of consecutive non-zero bins. Avalanche duration was determined by the length of this sequence, while avalanche size was measured as the total spike count within sequence. The analysis of neural avalanches and criticality is the same as the above experimental data.

**Estimate the effective connectivity from the neural activity**

In generally, inferring the underlying the network structure/effective connectivity from the activity is quite challenge, and there are some different well-established methods [55]. We here adopt a simple solution by assuming the linearity of dynamics where neural activity $x$ operates near the noise-driven steady state, and each node is driven by the independent Gaussian white noise with unit variance and the reciprocal structural interaction with other connected nodes. These assumptions give rise to the following equation of the dynamics:

$$\begin{cases} \dfrac{dx}{dt} = Ax + \xi \\ A = A^T \end{cases}.$$

The covariance matrix of the independent white noise driven activity $P = xx^T$ follows the Lyapunov equation $AP + PA^T = I$ [56]. Under the symmetry assumption of the structural/effective connectivity matrix A, we obtain $A = P^{-1}/2$ . For both experimental data and balanced model, we used the covariance matrix of neural activity to approximate $P$, and then solved the estimated structural/effective connectivity matrix $A$.

**Measurement of representation similarity and representation geometry**

Following the previous studies [9,36], we examined how the relationship between representations of multiple signals change with the time. We label the population response vector of each signal across different trial as $\{v_i^d \in R^{N_E}\}_{i=1,2\ldots M; d=1,2\ldots D'}$, and $M$ represents the number of signals and $D'$ represents the number of sessions. The representation similarity on each session is given by matrix $S_{ij}^d = \mathrm{corr}(v_i^d, v_j^d)$, and the correlation between the representation similarity matrices across $\Delta d$ session in Fig 3E is given by the $\langle \mathrm{corr}\,(\mathrm{vec}(S^d), \mathrm{vec}(S^{d+\Delta d}))\rangle_d$.

In the representational space spanned by the population response vectors where each signal is represented by a node, the geometric relationship among the signals are represented by the edges between the nodes, namely, the difference between each population response vector pair of the signals, $v_i^d - v_j^d$(edge between the signal $i$ and signal $j$ in Fig. 2F) in a given session $d$. If the geometric relationship is stable across sessions, then the angle $\theta$ between any two edges will be preserved. The angle (with unit in degree) between two edges $ij$ and $ik$ (see illustration in Fig. 2F) is calculated as $\theta_{ij,ik}^d = \frac{180}{\pi} \arccos \frac{(v_i^d - v_j^d)\cdot(v_i^d - v_k^d)}{\|v_i^d - v_j^d\|_2 \|v_i^d - v_k^d\|_2}$. The overall representational geometry is denoted by the vector consisting of all angles as $\boldsymbol{\theta}^d = \{\theta_{ij,ik}^d\}$, and the distance between the representational geometry across $\Delta d$ sessions in Fig 2G is given by the $\langle \|\theta^d - \theta^{d+\Delta d}\|_2 \rangle_d$.

**Linear discrimination analysis**

To evaluate low dimension neural representation, we employed Linear Discriminant Analysis (LDA). In the model simulation, our mean population response matrix $A \in R^{N_E \times T}$ was associated with signal labels $L \in Z^{1\times T}$, where $L_i \in [1, M]$, and $M$ represents the number of distinct stimuli or classes ($M$=10 in our study). Here $N_E$ represent the number of excitatory neurons, and $T = T_0 * M$, where $T_0$ represents the trial number of each stimulus ($T_0$=100 in our study). The fundamental idea of the LDA is to addresses the eigenvalue problem of $S_W^{-1} S_B$, where $S_W$ and $S_B \in R^{N_E \times N_E}$ represent the averaged within-class and between-class co-variance matrices, respectively. Here we utilized the coefficient matrix given by $S_W^{-1}\mu \in R^{N_E \times M}$, where $\mu = [\mu_1 - \bar{\mu}, \ldots, \mu_M - \bar{\mu}] \in R^{N_E \times M}$ with $\mu_m$ represents the mean activity vector for class m. and $\bar{\mu}$ represent the mean activity vector of all labels. This coefficient matrix corresponds to the linear combination of eigenvectors of $S_W^{-1} S_B$ and provides subspaces that effectively separate each class from others. For enhanced visualization (Fig. 2C, Fig. 4C, and Fig. 6C), we implemented t-SNE dimensionality reduction to obtain low-dimensional representations of the learned activity patterns $(S_W^{-1}\mu)^{-1} A$. We evaluated our LDA performance using 5-fold cross-validation to ensure robust assessment of its generalization capabilities.

Intuitively, we can regard the LDA as finding a learning subspace $(S_W^{-1}\mu)^{-1}A$ to decode the mean population response matrix $A$, and the $(S_W^{-1}\mu)^{-1}$ can be understood as the synapse strength from the excitatory neurons to the downstream decoding neural population. Since excitatory neurons only project excitatory synapses with positive strength, we incorporated a non-negative constraint on the learned coefficients using non-negative linear regression. This involved solving:

$$\mathrm{argmin}_{\mathrm{V}} \|VA - (S_W^{-1}\mu)^{-1}A\|_2^2, \qquad \text{subject to } V_{ij} \geq 0, \tag{7}$$

which effectively replace the original coefficient matrix by a non-negative one. While this non-negative LDA performed worse than standard LDA on training data, it demonstrated significantly superior generalization on test data. Then, similar LDA method is also used to analyze the experimental data (Fig. 6C).

**Acknowledgement:** We thank Prof. Daqing Guo and Prof. Pulin Gong for providing critical comments to our study.

**Funding:** This work was supported in part by STI 884 2030-Major Projects (2022ZD0208500) and the Hong Kong Research Grant Council (CRF C4012-22G, GRF 12201421).

**Author contributions**: Conceptualization: Z. Yang and C. Zhou. Methodology: Z. Yang, J. Liang and C. Zhou. Investigation: Z. Yang and C. Zhou. Visualization: Z. Yang, J. Liang, W Yung and C. Zhou. Supervision: C. Zhou. Writing—original draft: Z. Yang. Writing—review and editing: Z. Yang, J. Liang, W Yung and C.

Zhou. Resources: C. Zhou; Data curation: Z. Yang. Validation: Z. Yang. Formal analysis: Z. Yang. Software: Z. Yang. Project administration: C. Zhou. Funding acquisition: W Yung and C. Zhou.

**Competing interests**: the authors declare that they have no competing interests.

**Data and materials availability:** The original experimental data is sourced from https://portal.brain-map.org/circuits-behavior/visual-coding-2p. The program for analyzing the experimental data and performing numerical simulations of the model will soon be available at https://github.com/hkbucns.

## SI Figures

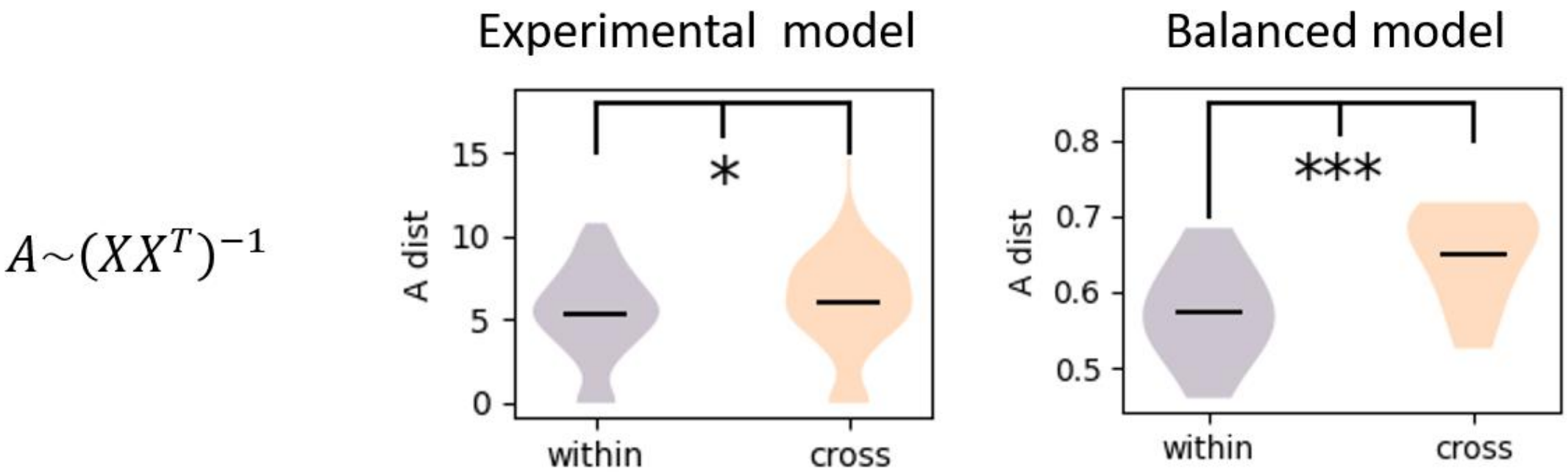

**Figure S1. The estimated effective connectivity from neural activity using inverse of co-variance matrix.** Both experimental data (two sample t test, $p \approx 0.03 < 0.2$) and balanced model (two sample t test, $p \approx 0$) show larger change in effective connectivity for the cross-session than the within-session.

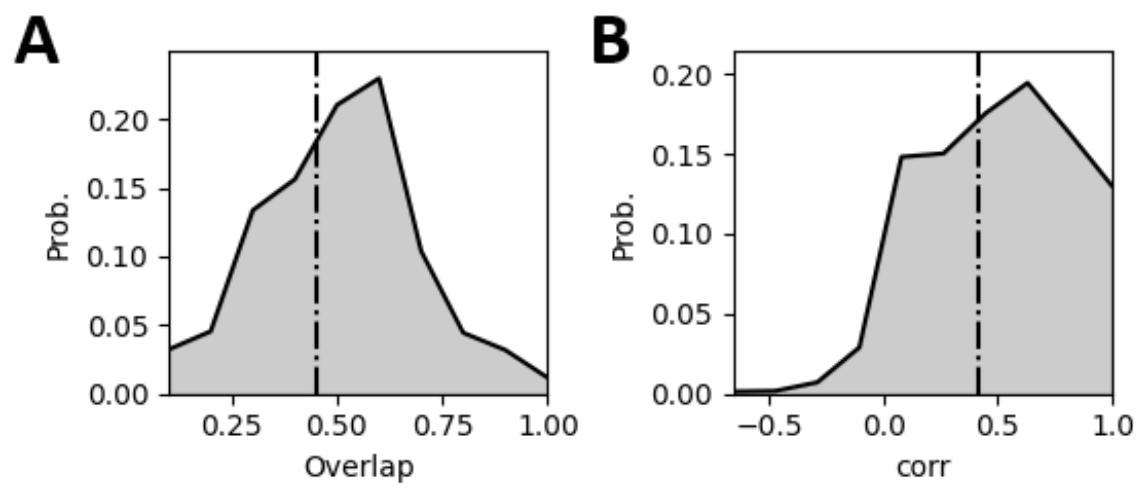

**Figure S2. The statistics of representational drift of the experimental data.** The overlap refers to the overlap ratio of the active neurons under the same signal across different sessions, and the correlation refers to the correlation between the population response vectors under the same signal across different sessions. The statistics of response pattern is based on $78$ mice and 30 signals.

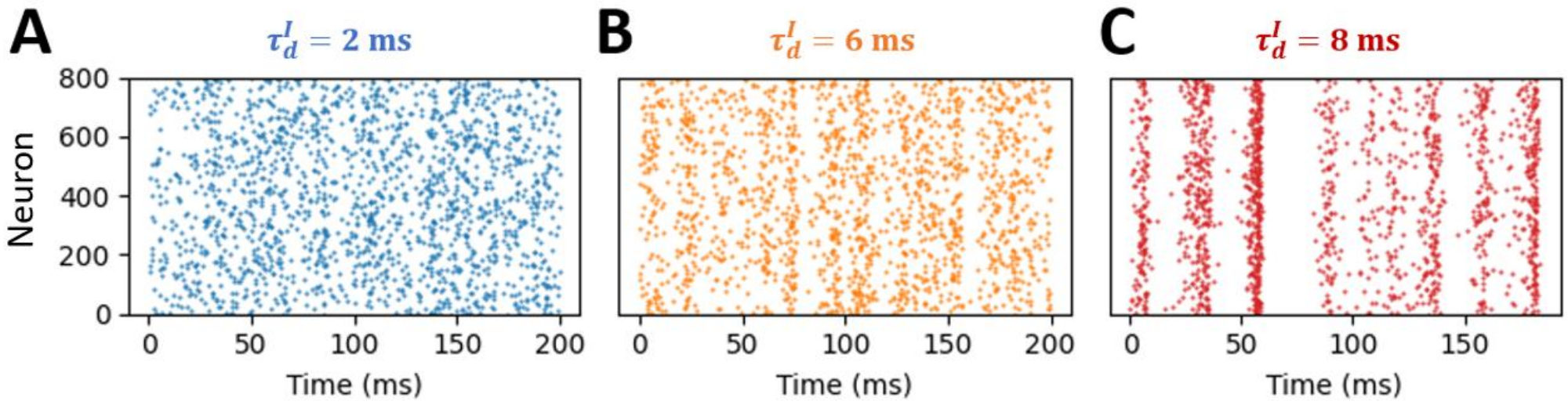

**Figure S3. Raster plots of plasticity model in different state.** The blue, orange, and red raster plots correspond to spikes of the excitatory neurons in the subcritical ($\tau_d^I = 2\ ms$), critical state ($\tau_d^I = 6\ ms$), and supercritical state ($\tau_d^I = 8\ ms$), respectively.

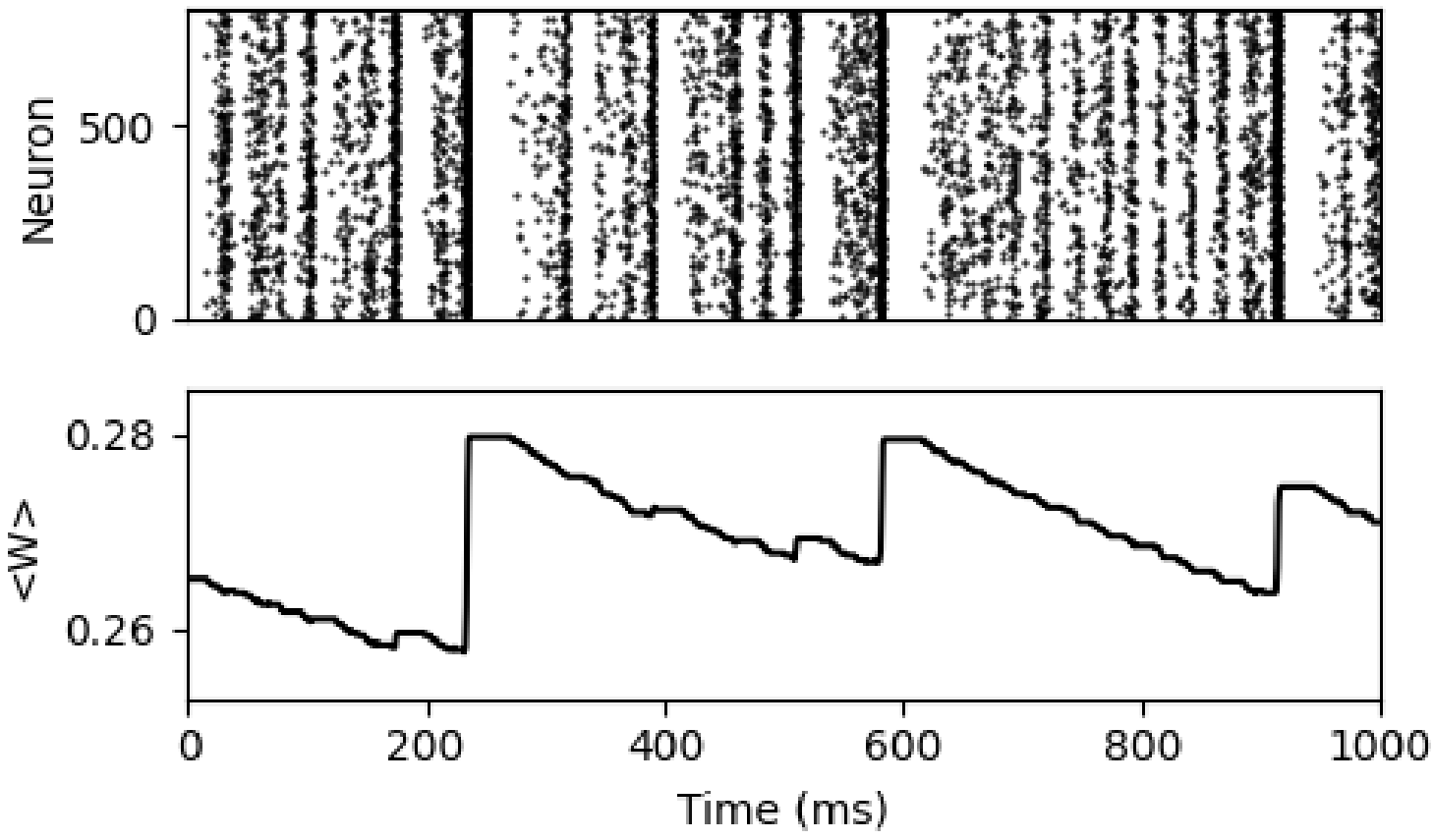

**Figure S4. Raster plot of the plasticity model at the supercritical state with $\tau_d^I = 10$ ms and the corresponding dynamics of the average synapse strength.**

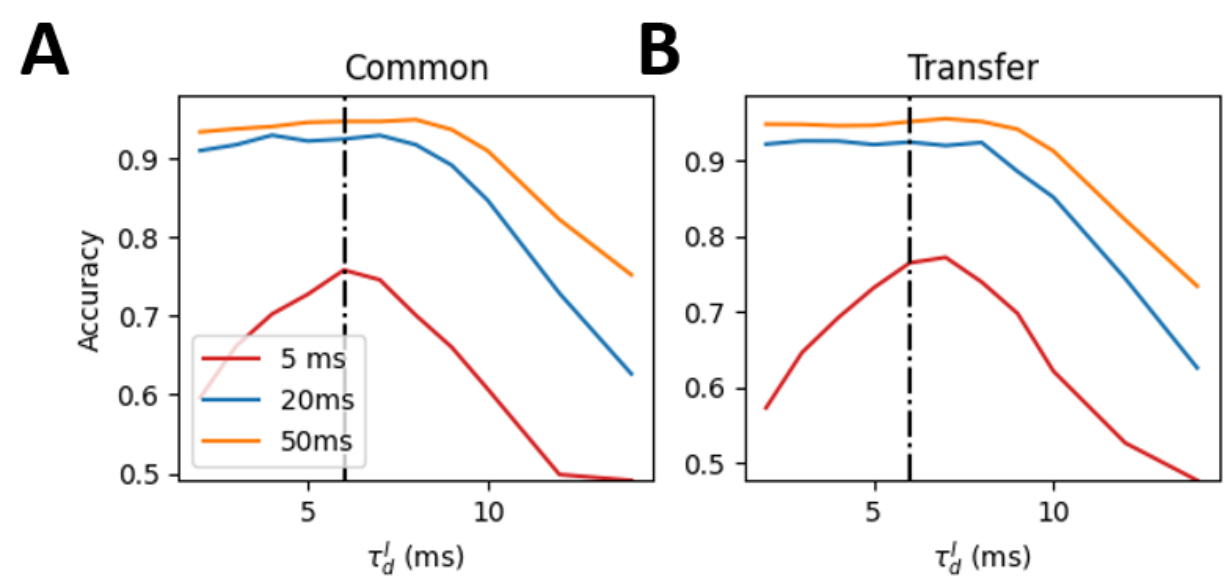

**Figure S5. Accuracy of different decoders with different input durations.** We use two decoders to evaluate the reliability of signal representation across sessions. For common decoder, we put the neural response from different sessions together, and used the LDA analysis to obtain the decoder from the training set, then evaluated the accuracy on the test set. For the transfer decoder, we used the neural response in one session to obtain the LDA-based decoder, and examined the decoding accuracy on the neural response in another session. The different color refers to different signal durations, and vertical black dash line indicates the critical state.

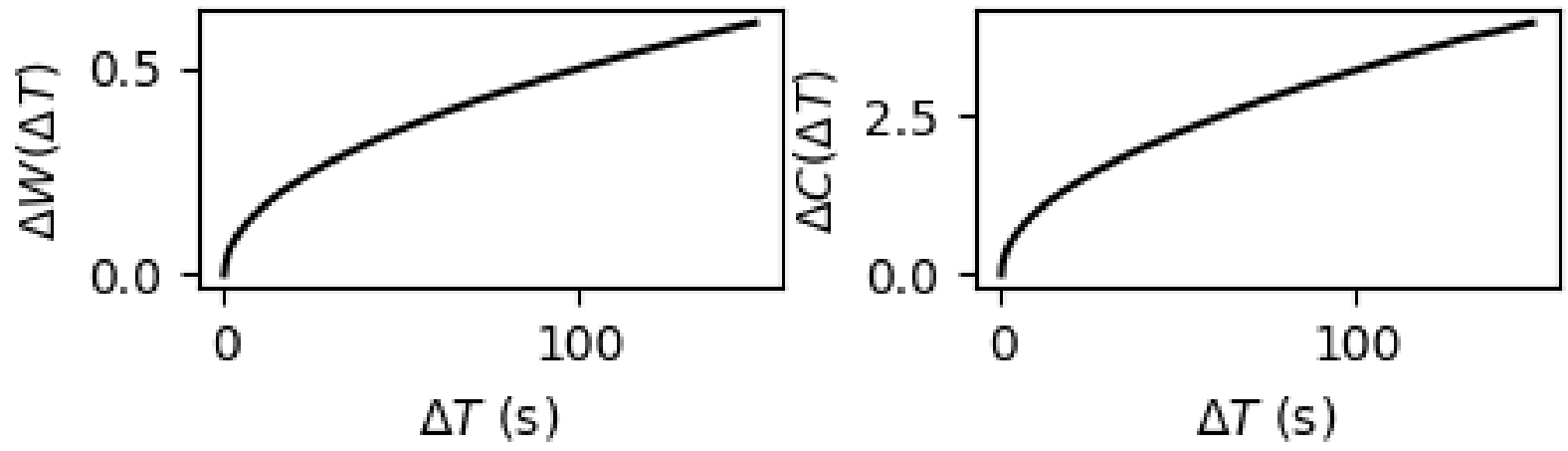

**Figure S6. The synapse strength and indegree patterns of the shuffled plasticity model.** We use the same shuffled plasticity model as the Fig. 5. Since the $\Delta W$ in each step follows the random distribution, then the distance of synapse strength and indegree patterns naturally increase with the $\Delta T$.